\documentstyle[aaspp4,epsf,equation,astrobib,11pt]{article}

\input{epsf}

\let\oldfootsep=\footnotesep
\def\msun { \rm {{\em M}_\odot}}

\def\t0{t_{\rm 0}}

\def\spose#1{\hbox to 0pt{#1\hss}}
\def\simlt{\mathrel{\spose{\lower 3pt\hbox{$\mathchar"218$}}
     \raise 2.0pt\hbox{$\mathchar"13C$}}}
\def\simgt{\mathrel{\spose{\lower 3pt\hbox{$\mathchar"218$}}
     \raise 2.0pt\hbox{$\mathchar"13E$}}}

\begin{document}

\title{Red Clump Stars as a Tracer of Microlensing Optical Depth}
\author{David P.~Bennett}      % bennett@nd.edu}
\affil{Department of Physics, University of Notre Dame, 
                 Notre Dame, IN 46556}
\affil{email: bennett@nd.edu}

% \newpage

\vspace{-5mm}
\begin{abstract} 
\rightskip = 0.0in plus 1em

Zaritsky and Lin have recently suggested that the color magnitude
diagram of the Large Magellanic Cloud (LMC) contains evidence of
foreground red clump stars. They interpret
this as evidence of tidal debris or a dwarf galaxy at a distance of
$\sim 35$ kpc which may be responsible for the large gravitational
microlensing optical depth observed by the MACHO Collaboration.
I derive a relationship between the microlensing optical depth of
such a foreground population and the observed density of
foreground red clump stars.  Recent observational determinations
for Pop I and Pop II stellar mass functions are used to show that
the surface density of foreground red clump stars claimed by Zaritsky and Lin
implies a microlensing optical depth in the range 
$\tau_{fg} = 0.8-3.6\times 10^{-8}$ which is only 3-13\% of 
$\tau_{LMC}$ as determined by the MACHO Collaboration. If the foreground
population has a similar star formation history to the LMC, then the implied
$\tau_{fg}$ is only 3-4\% of $\tau_{LMC}$.

\end{abstract}
\vspace{-5mm}

\newpage
\section{Introduction}
\label{sec-intro}

The MACHO Collaboration has recently reported a microlensing optical 
depth of $\tau_{LMC} = 2.9{+1.4\atop -0.9}\times 10^{-7}$ toward the Large
Magellanic Cloud (LMC)
\cite{macho_lmc2}. This is a substantial fraction of the microlensing
optical depth predicted by a standard Galactic model with a dark halo
composed entirely of massive compact objects (Machos), and it suggests
that at least a partial solution to the dark matter problem has been
found. However, the timescales of the detected microlensing
events seem to indicate a typical lens mass of $\sim 0.5\msun$ which is
in the mass range of normal main sequence stars. Main sequence stars
are too bright to comprise a significant amount of the Galaxy's dark
halo, so if the lensing objects are indeed composed of baryons, they
must be in another form. White dwarfs are an obvious possibility.
Since white dwarfs generally form as remnants of luminous stars, 
however, there
are a number of observational constraints on white dwarf halo models 
\cite{adams,chab,fields,gibmould}.  Another
possibility is that the lensing objects are not made up of  baryons
but are black holes which formed at the QCD phase transition
\cite{pbh}.

It is also possible that the microlensing events detected by the MACHO
collaboration do not represent a large contribution to the total mass
of the Galactic halo. \citeN{zhao_fd} has suggested
that there might be a previously unknown dwarf galaxy which,
by chance, happens to lie in the foreground of the LMC. It is
{\it a priori} rather unlikely for such a dwarf galaxy to be
positioned in front of the LMC, but this is still a possibility.
\citeN{zhao_tt} has also suggested that the foreground
object might be the debris of a dwarf galaxy that is in
the process of being tidally disrupted or perhaps even a tidal
tail of the LMC itself.

\citeN{zaritskylin} have
recently suggested that there may be evidence of such a foreground
population visible in color magnitude diagrams of stars seen in
the direction of the LMC. They find a feature on the giant branch
of the LMC color magnitude diagram which they interpret as 
a population of  red clump stars in the foreground of the LMC
at a distance of $\sim 35$ kpc.
They suggest that this population may be associated with a dwarf galaxy
or tidal tail that is massive enough to explain the large microlensing
optical depth observed by the MACHO Collaboration.

In this {\it Letter}, I address the implications of Zaritsky \& Lin's 
interpretation of their observations directly by
deriving a relation between the observed density of
red clump stars and the microlensing optical depth of a normal
stellar population that the red clump stars belong to. The normal
stellar population is assumed to have an initial mass function (IMF) 
similar to the observed mass function in the Galactic disk or in
globular clusters. It is then shown that the surface density of 
foreground red clump stars implied by Zaritsky \& Lin yields a microlensing
optical depth much smaller than the value observed by the MACHO Collaboration.

\section{Optical Depth of a Foreground Population}
\label{sec-od}

Let us assume that the foreground stellar population has a Salpeter-like
initial mass function: $ n(m) \propto m^{-2.3}$ for $m_1 < m < m_2$. Here
and throughout this paper lower case $m$ will always refer to a mass in
units of $\msun$. For $m < m_1$, $n(m)$
is assumed to be constant, and stars with $m > m_2 = 10$ are ignored.
This form of the IMF can be used with $m_1 = 0.6$ to represent a stellar
population similar to the Galactic disk (Pop I) \cite{diskimf}
and with $m_1 = 0.3$ to represent an older globular cluster type IMF
(Pop II) \cite{gcimf}.
For simplicity, the stellar population is described by a single turn-off
mass, $M_{to} = m_{to}\msun$, and a main-sequence lifetime
$t_{ms} = 1.1\times 10^{10} m_{to}^{-3.75}\,{\rm yrs}$ which is appropriate
for stars with $M \sim \msun$ \cite{milbin}. (The value quoted for
$t_{ms}$ is actually the pre-horizontal branch lifetime.) We will also
need the lifetime of the stars in the red clump. This
we take to be $t_{RC} = 10^8\,{\rm yrs}$ independent of the star's initial
mass \cite{redclump}.
Finally, we must consider the fate of the stars which have evolved past the 
horizontal branch. These are assumed to have become white dwarfs with
a mass given by $M_{wd} = 0.15M + 0.38\msun$ \cite{wdmass}.

With these assumptions it is straight forward to work out the expected
microlensing optical depth for a foreground stellar population with
a measured surface density of red clump stars. The total mass in stars
is given by
\begin{equation}
\begin{eqalign}
\label{m-all}
{\cal M} =& A\left[ \int_0^{m_1} m_1^{-1.3} dm
                  + \int_{m_1}^{m_{to}} m^{-1.3} dm
                  + \int_{m_{to}}^{m_2} m^{-2.3} m_{wd} dm \right] \\
         =& A\left[ 4.333m_1^{-0.3} - 0.2653 - 2.833 m_{to}^{-0.3}
                    + 0.2923 m_{to}^{-1.3} \right] \ ,
\end{eqalign}
\end{equation}
where $A$ is an arbitrary normalization constant.
In order to determine the total number of red clump stars we must determine
the interval in initial mass that the red clump progenitor span. This is
just the interval $m$ to $m(1+\delta)$ where $\delta = t_{RC}/(3.75 t_{ms})$
since $t_{ms} \propto m^{-3.75}$. Thus, the total number of red clump stars
is given by
\begin{equation}
\begin{eqalign}
\label{n-rc}
 N_{RC} =& A \int_{m_{to}}^{(1+\delta)m_{to}} m^{-2.3} dm \\
        =& A\, m_{to}^{-1.3}\delta = 0.00242\, A\, m_{to}^{2.45} \ .
\end{eqalign}
\end{equation}
Equations (\ref{m-all}) and (\ref{n-rc}) give the total stellar mass per red
clump star:
\begin{equation}
\label{mperc}
  {{\cal M}\over N_{RC}} = m_{to}^{-2.45} \left( 1788m_1^{-0.3} - 110 
                  - 1169m_{to}^{-0.3} + 121m_{to}^{-1.3} \right) \ .
\end{equation}
Now, the microlensing optical depth of a foreground population 
is just the fraction of the sky covered by the Einstein rings of all
the lenses.  The linear size of the Einstein ring
radius is given by $R_E = 2\sqrt{GMx(1-x)L}$ for a lens of mass $M$ at
a distance of $xL$ for source stars at a distance $L$. Thus, the microlensing
optical depth of a foreground population with an angular surface density of
$1\msun$ per square degree is just
\begin{equation}
\label{tau-1}
   \tau_1 = {1-x\over x} 3.95\times 10^{-14} \ ,
\end{equation}
where we have assumed source stars in the LMC at 50 kpc. Equations 
(\ref{mperc}) and (\ref{tau-1}) can be combined to yield an expression
for the microlensing optical depth of a foreground population of stars
with $\sigma_{RC}$ red clump stars per square degree:
\begin{equation}
\label{tau-fg}
   \tau_{fg} = 3.95\times 10^{-14}{1-x\over x}{\sigma_{RC}\over m_{to}^{2.45}}
     \left( 1788m_1^{-0.3} - 110 - 1169m_{to}^{-0.3} + 121m_{to}^{-1.3} \right)
       \ .
\end{equation}
A couple of trends are apparent from equation (\ref{tau-fg}). First, a Pop I
initial mass function ($m_1 = 0.6$) implies a higher microlensing optical
depth than a Pop II IMF ($m_1 = 0.3$) for a fixed turn-off mass. This is
because the Pop II IMF has more low mass stars per red clump star. Also, for
a fixed IMF, an old population with a lower turn-off mass will have a
higher microlensing optical depth per red clump star. This is because older
stars spend a smaller fraction of their lifetime as red clump stars.

\section{Application to Zaritsky \& Lin's Observations}
\label{sec-zl}

\begin{figure}
\plotone{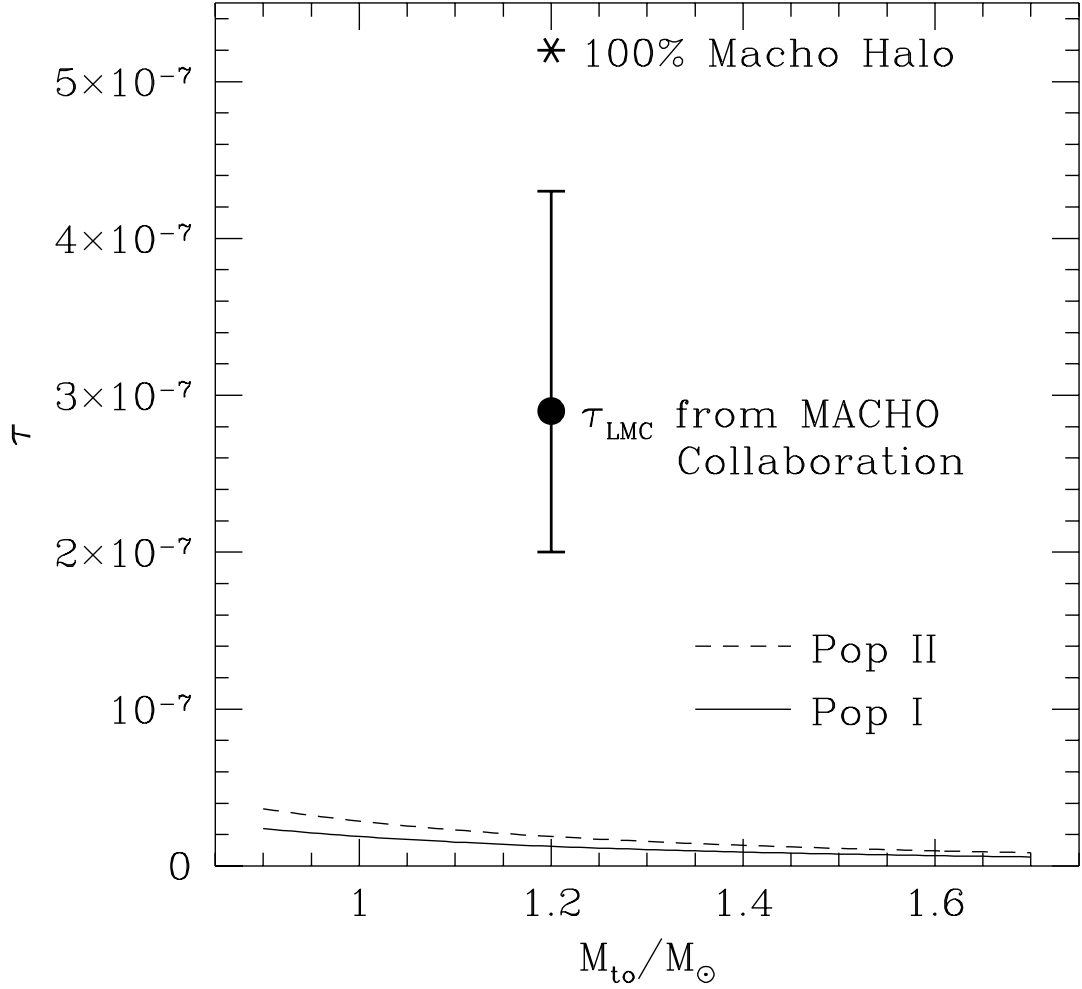}
\caption{The microlensing optical depth of a population of stars
with a angular density of $\sigma_{RC} = 1200$ red clump stars per square
degree located at 35 kpc in the foreground of the LMC is plotted as a function
of the main sequence turn-off mass, $M_{to}$. The solid line assumes a
Pop I initial mass function and the dashed line assumes a Pop II initial
mass function. For comparison, the microlensing optical depth, $\tau_{LMC}$,
measured by the MACHO Collaboration and the predicted optical depth for
a standard halo composed entirely of Machos are also shown. (Strictly speaking,
$\tau_{LMC}$ is a lower limit on the microlensing optical depth because
MACHO is only sensitive to microlensing events with Eintein diameter 
crossing times between 2 and 200 days.)
\label{fig}}
\end{figure}

Equation (\ref{tau-fg})
can now be used to estimate the microlensing optical depth of the
population of foreground stars that \citeN{zaritskylin} have
inferred from their LMC color magnitude diagram. They 
have identified a feature in their color magnitude diagram which
they interpret as a red clump population in the foreground of the LMC
at a distance of $\approx 35\,$kpc, but other interpretations of these
observations are certainly possible. It could be that this 
feature on the LMC giant branch is caused by stellar evolution, for example.
For the purpose of this paper, however, we will assume that Zaritsky \& Lin's
interpretation of their observations is correct.
The ``foreground" clump population
is estimated to have a surface density of $\simlt 5$ to 7\% of the 
surface density of LMC red clump stars in the fields that they observed.
They found approximately 70,000 stars in the red clump and the giant
branches in the three square degrees of their survey which implies
that the foreground population should have approximately $\sigma_{RC} = 1200$
red clump stars per square degree. Inserting this value and $x=0.7$
(for a population at 35 kpc) into equation (\ref{tau-fg}) yields the
results shown in Figure \ref{fig}. This figure shows the estimated optical
depth as a function of the main sequence turn-off mass for both
Pop I ($m_1 = 0.6$) and Pop II ($m_1 = 0.3$) initial mass functions.
The models for stellar populations in the foreground of the LMC
which seem the most plausible hold that the
foreground object is composed of tidal debris from either LMC-Milky Way
interactions or LMC-SMC interactions \cite{zhao_tt,zaritskylin}.
These models would suggest that the foreground population is relatively
young, like the LMC. Thus, $m_{to}=1.5$ and $m_1=0.6$ are probably appropriate
for such a population. Inserting these numbers in equation (\ref{tau-fg})
yields a foreground microlensing optical depth of 
$\tau_{fg} = 7.6\times 10^{-9}$ which is 2.6\% of $\tau_{LMC}$ as observed by
MACHO. Thus, it is unlikely that tidal debris in the foreground of
the LMC can explain a significant amount of the observed microlensing
optical depth.

A larger microlensing optical depth can be obtained if we consider the
extreme, old Pop II case with $m_{to}=0.9$ and $m_1=0.3$. This yields a
foreground microlensing optical depth of $\tau_{fg}=3.6\times 10^{-8}$, but
this is still only 13\% of the observed value. Furthermore, an old population
should have a large number of RR Lyrae stars which could be seen in the
foreground of the LMC. Such foreground RR Lyrae are not seen in the MACHO
database \cite{macho_newd} although the search of the MACHO database was
limited to RR Lyrae brighter than $V = 18$ which means that they are only
sensitive relatively bright RR Lyrae at 35 kpc.

\section{Discussion and Conclusions}
\label{sec-conclude}

Using initial mass functions determined from recent HST observations of
globular clusters and the Galactic disk, I have derived a relationship
between the surface density of red clump stars and the microlensing
optical depth of the parent stellar population. This relationship has
been used to show that the foreground population suggested by
Zaritsky \& Lin does not have a mass in stars to explain the microlensing
optical depth observed toward the LMC by the MACHO Collaboration.

A number of other authors have also estimated the microlensing optical
depth for such a foreground stellar population. Zaritsky \& Lin estimated
$\tau_{fg} = 1.2\times 10^{-7}$ which is 40\% of $\tau_{LMC}$ as measured
by the MACHO Collaboration and 3-16 times the estimates derived here.
Their estimate is based upon an estimate of the surface density of the
LMC disk which assumes that all of the mass of the central 3 kpc of the
LMC is made up of stars in the disk. \citeN{gould_fg} points out that 
this is an unreasonable assumption for the LMC, and he notes that Zaritsky \& 
Lin made an error when correcting the apparent LMC surface density for
inclination. Zaritsky \& Lin also use the
spherical formula to estimate the mass of the LMC disk (which is not very
spherical) and make no correction for the fact that their
field is a low density ``inter-arm" region of the LMC disk (Freeman,
private communication). Each of these errors has an effect
at about the factor of 2 level, but all of them tend to make 
Zaritsky \& Lin's $\tau_{fg}$ estimate too large. The combination of
these errors can easily explain the factor of 3-16 discrepancy with the
results presented here.

\citeN{gould_fg} and \citeN{johnston_fg} have also considered the
microlensing optical depth of a foreground object like that proposed
by Zaritsky \& Lin. Gould has pointed out that the surface brightness
profile of the LMC does not allow for a foreground stellar population
with a substantial microlensing optical depth to extend beyond 5 degrees
from the LMC center. Gould's argument would fail for a foreground
population that had a surface brightness distribution so uniform that
it could be confused with the sky brightness, however.
Johnston has shown theoretically that tidal debris
from the LMC or another galaxy would either extend well beyond the LMC
or have a very short lifetime unless the associated microlensing
optical depth was small. Together the Gould and Johnston arguments show
that the scenario in which a foreground of tidal debris accounts for the
microlensing seen toward the LMC is rather unlikely. The calculation
presented here strengthens their case, and it also applies to objects
such as dwarf galaxies that might not extend beyond the foreground of
the LMC. Thus, while it is not certain that Zaritsky \& Lin's
observations really indicate that a foreground stellar population exists,
there is now rather strong evidence that such a foreground population of
normal stars cannot account for the LMC microlensing events seen by
the MACHO Collaboration.

\acknowledgements
\section*{Acknowledgments}

I would like to thank Ken Freeman, Dante Minniti, and Dennis Zaritsky for 
useful discussions regarding the Zaritsky \& Lin data, and I'd like to
thank Stuart Marshall, Dante Minniti,
Sun Hong Rhie, and Chris Stubbs for comments on an early draft of
this paper.  This work was supported in part by the NSF through the
Center for Particle Astrophysics.

%%%\clearpage

%%%\section*{References}
% \bibliography{apjmnemonic,general_refs,macho_refs,binary_refs}
\bibliographystyle{apj}

\end{document}